\documentclass[twocolumn]{aastex701}

\usepackage{CJK}
\usepackage{xspace}
\usepackage{hyperref}

\newcommand{\Oumuamua}{\okina Oumuamua\xspace}

\newcommand{\Nitau}{$3880\pm39$\xspace}

\newcommand{\CNtau}{$6053\pm68$\xspace}
\newcommand{\Ctwotau}{$4194\pm45$\xspace}
\newcommand{\Cthreetau}{$3833\pm45$\xspace}

\hypersetup{pdfauthor={Hoogendam}}
\begin{document}

\DeclareRobustCommand{\okina}{%
  \raisebox{\dimexpr\fontcharht\font`A-\height}{%
    \scalebox{0.8}{`}%
  }%
}

\title{Post-Perihelion Integral Field Spectroscopy of the Interstellar Comet 3I/ATLAS}

\author[0000-0003-3953-9532]{Willem~B.~Hoogendam}
\altaffiliation{NSF Graduate Research Fellow}
\affiliation{Institute for Astronomy, University of Hawai\okina i, 
2680 Woodlawn Drive, Honolulu, HI 96822, USA}
\email{willemh@hawaii.edu}  

\author[0000-0002-6230-0151]{David~O.~Jones}
\affiliation{Institute for Astronomy, University of Hawai\okina i, 640 N. A\okina ohoku Pl., Hilo, HI 96720, USA}
\email{dojones@hawaii.edu}

\author[0000-0002-5033-9593]{Bin~Yang}
\affiliation{Instituto de Estudios Astrof\'isicos, Facultad de Ingenier\'ia y Ciencias, Universidad Diego Portales, Santiago, Chile}
\affiliation{Planetary Science Institute, 1700 E Fort Lowell Rd STE 106, Tucson, AZ 85719, USA}
\email{bin.yang@mail.udp.cl} 

\author[0000-0003-4631-1149]{Benjamin~J.~Shappee}
\affiliation{Institute for Astronomy, University of Hawai\okina i, 
2680 Woodlawn Drive, Honolulu, HI 96822, USA}
\email{shappee@hawaii.edu} 

\author[0000-0001-5559-2179]{James~J.~Wray}
\affiliation{School of Earth and Atmospheric Sciences, Georgia Institute of Technology, 311 Ferst Drive, Atlanta, GA 30332, USA}
\affiliation{Institute for Astronomy, University of Hawai\okina i, 
2680 Woodlawn Drive, Honolulu, HI 96822, USA}
\email{jwray@gatech.edu}

\author[0000-0002-2058-5670]{Karen~J.~Meech}
\affiliation{Institute for Astronomy, University of Hawai\okina i, 
2680 Woodlawn Drive, Honolulu, HI 96822, USA}
\email{meech@hawaii.edu}  

\author[0000-0002-5221-7557]{Christopher~Ashall}
\affiliation{Institute for Astronomy, University of Hawai\okina i, 
2680 Woodlawn Drive, Honolulu, HI 96822, USA}
\email{cashall@hawaii.edu}

\author[0000-0002-2164-859X]{Dhvanil~D.~Desai}
\affiliation{Institute for Astronomy, University of Hawai\okina i, 
2680 Woodlawn Drive, Honolulu, HI 96822, USA}
\email{dddesai@hawaii.edu}

\author[0000-0001-9668-2920]{Jason~T.~Hinkle}
\altaffiliation{NHFP Einstein Fellow}
\affiliation{Department of Astronomy, University of Illinois Urbana-Champaign, 1002 West Green Street, Urbana, IL 61801, USA}
\affiliation{NSF-Simons AI Institute for the Sky (SkAI), 172 E. Chestnut St., Chicago, IL 60611, USA}
\affiliation{Institute for Astronomy, University of Hawai\okina i, 
2680 Woodlawn Drive, Honolulu, HI 96822, USA}
\email{jhinkle6@hawaii.edu}

\author[orcid=0000-0002-8732-6980]{Andrew~M.~Hoffman}
\affiliation{Institute for Astronomy, University of Hawai\okina i, 
2680 Woodlawn Drive, Honolulu, HI 96822, USA}
\email{amho@hawaii.edu}  

\author[0000-0001-7186-105X]{Kyle~Medler}
\affiliation{Institute for Astronomy, University of Hawai\okina i, 
2680 Woodlawn Drive, Honolulu, HI 96822, USA}
\email{kmedler@hawaii.edu}

\author[0000-0002-7305-8321]{Cameron~Pfeffer}
\affiliation{Institute for Astronomy, University of Hawai\okina i, 
2680 Woodlawn Drive, Honolulu, HI 96822, USA}
\email{cpfeffer@hawaii.edu}

\author[0000-0003-4936-4959]{Ruining~Zhao}
\affiliation{National Astronomical Observatories, Chinese Academy of Sciences, Beijing 100101, China}
\email{rnzhao@nao.cas.cn} 

\begin{abstract}
The environs of other stellar systems may be directly probed by analyzing the cometary activity of interstellar objects. The recently discovered interstellar object 3I/ATLAS was the subject of an intensive worldwide follow-up campaign in its pre-perihelion approach. Now, 3I/ATLAS has begun its post-perihelion departure from the Solar System. In this letter, we report the first post-perihelion blue-sensitive integral-field unit spectroscopy of 3I/ATLAS using the Keck Cosmic Web Imager on November 16, 2025. We confirm previously reported CN, Fe, and Ni outgassing along with detections of carbon chain molecules $\mathrm{C}_2$ and $\mathrm{C}_3$. We calculate production rates for each species. We find Fe and Ni production rates of $\mathrm{Q_{Fe}} = (9.55\pm3.96)\times10^{25}$~atoms~s$^{-1}$, and
$\mathrm{Q_{Ni}} = (6.61\pm2.74)\times10^{25}$~atoms~s$^{-1}$, resulting in a ratio of $\log(\mathrm{Q_{Ni}} / \mathrm{Q_{Fe}}) = -0.16 ± 0.03$, which matches Solar System comets well and continues the pre-perihelion trend of declining $\log(\mathrm{Q_{Ni}} / \mathrm{Q_{Fe}})$ with $r_h$. We investigate the radial distributions of these elemental species and find characteristic $e$-folding radii of \Nitau~km for Ni, \CNtau~km for CN, \Ctwotau~km for $\mathrm{C}_2$, and \Cthreetau~km for $\mathrm{C}_3$. Compared to pre-perihelion measurements, these radii have increased by a factor of $\sim$6.5--7. Our post-perihelion observations reveal that 3I/ATLAS continues to exhibit cometary behavior broadly consistent with Solar System comets. 
\end{abstract}


\section{Introduction}\label{sec:intro}

\begin{figure*}
\includegraphics[width=\textwidth]{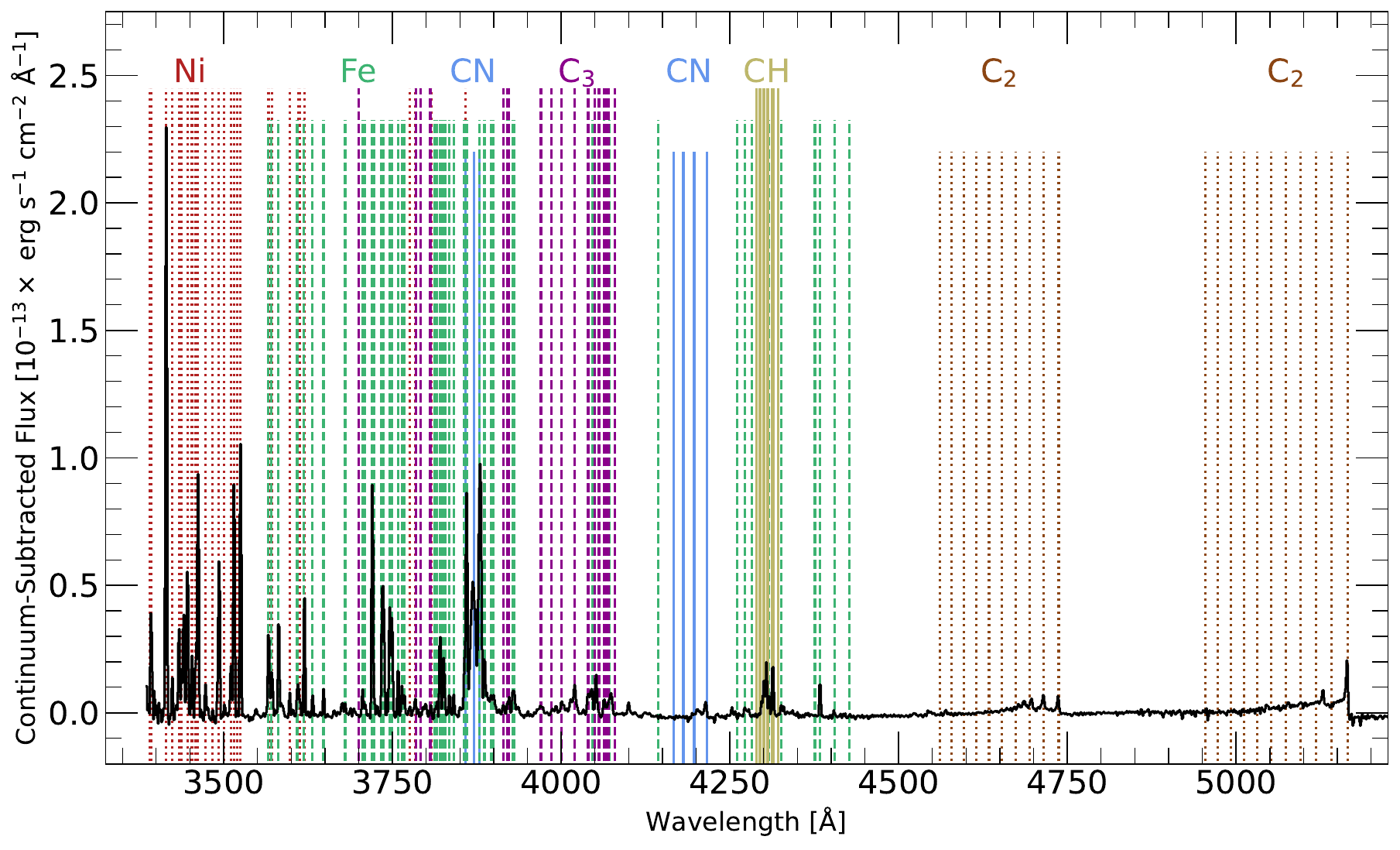}
\caption{The continuum-subtracted KCWI spectrum of 3I/ATLAS between 3325 \AA\ and 5225 \AA, extracted from a 2\arcsec\ aperture centered on the comet. Cometary emission species are denoted as follows: Ni as red dotted lines, Fe as green dashed lines, CN as solid blue lines, $\mathrm{C}_2$ as dotted light purple lines, and $\mathrm{C}_3$ as dashed dark purple lines. The $\mathrm{C}_2$ and $\mathrm{C}_3$ lines shown may not be individually resolved. }
\label{fig:spec1d}
\end{figure*}

Comets and asteroids ejected from other stellar systems are predicted to intercept the solar system by chance at a frequent rate \citep[e.g.,][]{Engelhardt2017}, yet the discovery of these objects has been limited to three exceptional examples: 1I/\Oumuamua\ \citep{Meech2017, ISSI_1I_review}, 2I/Borisov \citep{borisov_2I_cbet, Jewitt2019b, Guzik:2020}, and now, 3I/ATLAS \citep{Denneau2025, Seligman2025, Tonry2025}, which was recently discovered by the Asteroid Terrestrial-impact Last Alert System (ATLAS; \citealp{Tonry2018a, Tonry2025}). Just as small bodies in our Solar System reflect its primordial composition \citep[e.g.,][]{Bodewits2024}, interstellar objects are likewise probes of the composition and evolution of small bodies around different stars \citep[e.g.,][]{Jewitt2023ARAA, Fitzsimmons2024}.  

To date, each interstellar object has shown unique activity signatures. Despite observations throughout its brief window of visibility, a coma or outgassing activity was undetected for 1I/\Oumuamua \citep[e.g.,][]{Meech2017, Ye2017, Jewitt2017, ISSI_1I_review, Trilling2018}; however, outgassing was inferred from its non-gravitational acceleration \citep{Micheli2018}. Contrarily, 2I/Borisov was clearly outgassing, had a dusty coma \citep{Fitzsimmons:2019, Jewitt2019b, Cremonese2020, Guzik:2020, Hui2020, Kim2020, McKay2020, ye2020_borisov, yang2021}, and displayed many features commonly observed in Solar System comets \citep[e.g.,][]{Opitom:2019-borisov, Fitzsimmons:2019, Lin2020, McKay2020, Xing2020, Bannister2020}, including Ni and Fe outgassing \citep{Guzik2021, Opitom2021}.

3I/ATLAS is likewise undergoing cometary activity and has a visible coma \citep{Seligman2025, Jewitt2025, Cordiner2025, Rahatgaonkar2025, Opitom2025, delaFuenteMarcos2025, Chandler2025, Lisse2025, Hoogendam25_KCWI, Tonry2025}. The initial spectra exhibited red-sloped reflectance without strong emission features \citep{Seligman2025, Opitom2025, Puzia2025}. As it approached perihelion, outgassing increased. Absorption from large water ice grains in the coma \citep{Yang2025} and emission from CN \citep{Rahatgaonkar2025}, Ni \citep{Rahatgaonkar2025}, $\mathrm{CO}_2$ \citep{Lisse2025, Cordiner2025}, CO \citep{Cordiner2025}, and a likely extended source of OH emission \citep{Xing2025} has been reported. 

Integral field unit (IFU) data provide several advantages over slit spectroscopy, including the addition of spatial information. Previous blue-sensitive IFU data for 3I/ATLAS include those presented by \citet{Seligman2025} and \citet{Hoogendam25_SNIFS} from the SuperNova Integral Field Spectrograph (SNIFS; \citealp{Lantz2004}, see also \citealp{Tucker2022}) and Keck Cosmic Web Imager (KCWI; \citealp{Morrissey18}) presented by \citet{Hoogendam25_KCWI}. The latter revealed centrally concentrated Ni emission relative to CN with a less extended radial distribution. Here, we present the first post-perihelion IFU observations of 3I/ATLAS from KCWI.

\begin{figure*}
\includegraphics[width=\textwidth]{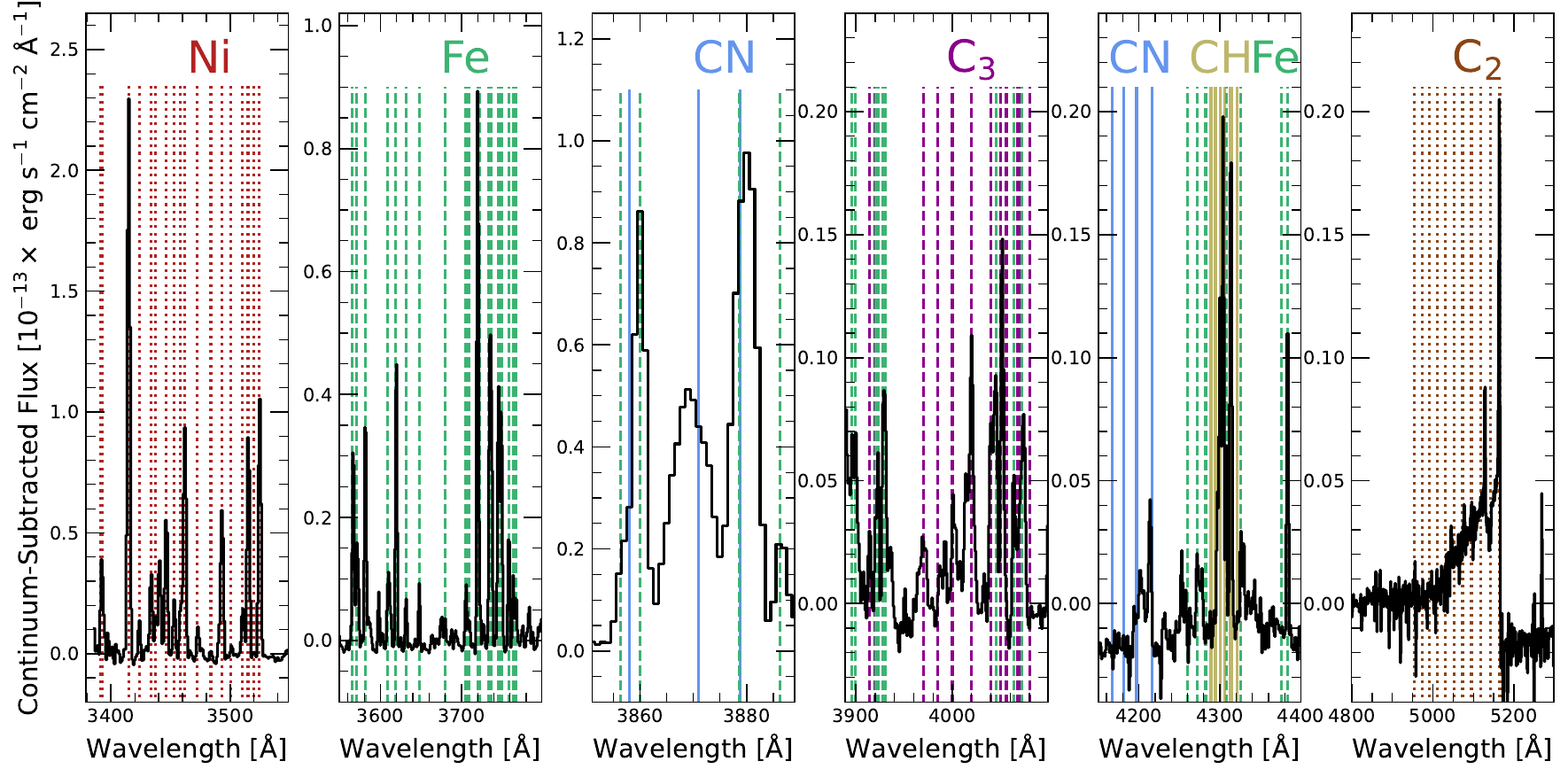}
\caption{The same as Figure~\ref{fig:spec1d}, but with individual panels for each emission feature.}
\label{fig:spec1d_zoom}
\end{figure*}

\section{Data}\label{sec:data}
We obtained a KCWI spectrum of 3I/ATLAS on UTC 2025-Nov-16 14:44:08 (observation start), along with two solar analogs (HD~103218 and HD~103390) and a flux calibration standard (Feige 67) at a similar airmass ($\sim$3, due to solar and lunar constraints). 3I/ATLAS was at heliocentric ($r_h$) and geocentric ($\Delta$) distances of 1.509~au and 2.089~au, respectively, at the start of the observations. The phase angle was 26.049\degr, and the true anomaly was 28.099\degr. 

Our KCWI configuration used the medium image slicer, which provides a 16.5\arcsec\ by 20.4\arcsec\ field of view with a slice width of 0.70\arcsec. The blue/red channel used the BL/RL gratings, respectively. These provide a spectral resolving power of $R\approx1800$ in the blue from $\sim$3300~\AA\ to $\sim$5500~\AA\ and $R\approx1000$ in the red from $\sim$5500~\AA\ to $\sim$9000~\AA\ in the red. The total on-source integration time was 300~s for the blue channel and 420~s for the red channel. Because 3I/ATLAS filled the entire medium slicer field of view, we also took separate sky frames near 3I/ATLAS in an empty field to enable accurate sky subtraction. Finally, we used a 2\arcsec\ aperture to extract 1D spectra from the cubes.

We reduced the data using the KCWI Data Reduction Pipeline \citep[DRP;][]{Neill2023_KCWIDRP} but disabled the pipeline's sky subtraction and flux calibration. Instead, we performed manual sky subtraction and flux calibration using the differential-atmospheric-refraction-corrected cubes. We extracted 1D spectra from the separate sky frames in the same manner as for the science observations, then median-combined them to produce a master sky spectrum, which was subtracted from the final target spectrum. We flux-calibrated the sky-subtracted frames using the CALSPEC reference spectrum of Feige 67 to compute a sensitivity function; the spectra were then multiplied by this function to produce flux-calibrated spectra. Lastly, we combined the spectra using inverse-variance weighting to produce final spectra for 3I/ATLAS and the two solar analog stars.

\section{Spectrospatial Analysis of 3I/ATLAS}\label{sec:analysis}

To subtract the solar continuum contribution, we use HD~203218 as a solar analog. We model the continuum in the same manner as \citet{Rahatgaonkar2025} and \citet{Hoogendam25_KCWI}. This model has a functional form of  
\begin{equation}
    F_{\mathrm{cont}}(\lambda) = R(\lambda)\times F_{\odot\mathrm{analog}}\left[\lambda\left(1+\frac{v}{c}\right)+\delta\lambda\right].
\end{equation}

\noindent In this model, the second-order polynomial ${R\left(\lambda\right)\equiv\frac{1}{S}\left(1+b_1\lambda+b_2\lambda^2\right)}$ models the cometary reflectance as a function of the the $b_1$ and $b_2$ parameters. The reflectance function is normalized by a factor $S$ to compensate for flux differences between the comet spectrum and the solar analog spectrum. Of note, 3I/ATLAS and the analog stars have a similar magnitude, making the normalization factor $S$ near unity in this instance. $F_{\odot\mathrm{analog}}$ is the solar analog flux, and is shifted by fitted nuisance parameters $v$ and $\delta\lambda$. 

Figure \ref{fig:spec1d} shows the 3325 \AA\ to 5225 \AA\ 1D continuum-subtracted spectrum from our KCWI datacube, and Figure \ref{fig:spec1d_zoom} shows the same continuum-subtracted spectrum over the wavelength ranges for each observed feature.  

\begin{figure}
\includegraphics[width=\linewidth]{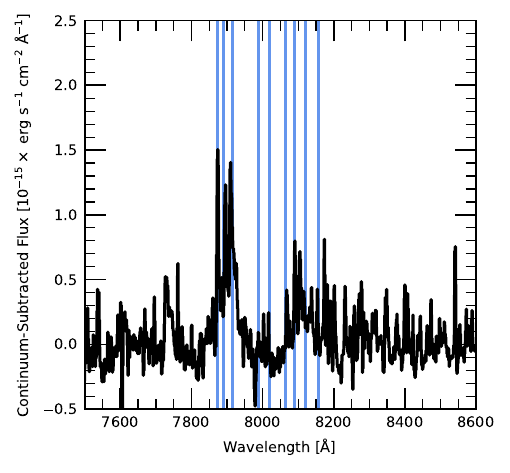}
\caption{The continuum-subtracted KCWI spectrum of 3I/ATLAS between 7500 \AA\ and 8600 \AA, extracted from a 3\arcsec\ aperture centered on the comet. The red CN system is shown in blue.}
\label{fig:spec1d_CN_red}
\end{figure}

\subsection{Activity Signatures}

Optical spectroscopy of 3I/ATLAS initially revealed pre-perihelion emission of Ni and CN \citep{Rahatgaonkar2025, Hoogendam25_KCWI, Hoogendam25_SNIFS, SalazarManzano25}, followed by detections of Fe \citep{Hutsemekers25}. Astronomer's Telegrams from \citet{Bolin25_ATel_C2C3}, \citet{Jehin25_ATel}, and \citet{Jehin25_ATel_2} report pre-perihelion detections of $\mathrm{C}_2$ and $\mathrm{C}_3$ and \citet{Ganesh25_ATel} and \citet{Bolin25_ATel_PostPeri} report the same in post-perihelion data.

We show post-perihelion detections of classic comet features such as $\mathrm{C}_2$ ($d^3\Pi_g - a^3\Pi_u$ (0,0) and (1,0) lines; \citealt{Swan_1857}), $\mathrm{C}_3$ ($A^1\Pi_u-X^1\Sigma_g^+$ lines; \citealp{Huggins1881, Gausset65}), and CH ($A^2\Delta-X^2\Pi$ (1,1) lines; \citealp{Meier98}), and find continued Ni and Fe activity, even after solar passage. 
Additionally, while \citet{Hoogendam25_SNIFS} and \citet{SalazarManzano25} only report detecting the CN violet system ($B^2\Sigma^{+} - X^2\Sigma^+$, $\Delta\nu = 0$) near 3870~\AA, we detect two additional CN sequences: the CN blue system ($B^2\Sigma^+ - X^2\Sigma^+$, $\Delta\nu = +1$) near $\sim$4200~\AA\ and the CN red system ($A^2\Pi - X^2\Sigma^+$, $\Delta\nu = +1$) near $\sim$8000~\AA\ (Figure \ref{fig:spec1d_CN_red}).

\begin{figure}
\includegraphics[width=\linewidth]{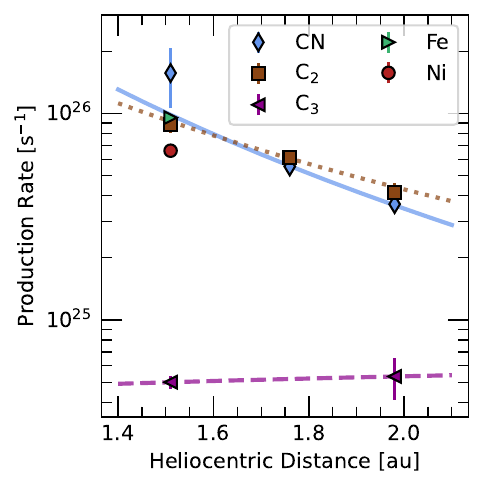}
\caption{Post-perihelion production rate evolution as a function of heliocentric distance using data from this work at $r_h=1.509$~au and rates reported from TRAPPIST measurements \citep{Jehin25_ATel, Jehin25_ATel_2}. CN and $\mathrm{C_2}$ have similar exponential decays, whereas $\mathrm{C_3}$ is nearly constant. Best-fit power laws are shown as the solid (CN), dotted ($\mathrm{C_2}$), and dashed ($\mathrm{C_3}$) lines.}
\label{fig:prod_rate_evo}
\end{figure}

We use a simple \citet{Haser:1957} model to convert the measured line fluxes into gas production rates for CN, $\mathrm{C_2}$, and $\mathrm{C_3}$. The number of photons emitted per molecule per second (i.e., the $g$-factor) and the scale lengths were taken from \citet{A'Hearn:1995}. The extracted spectrum has a physical radius of $\sim$3\,000~km. Our calculation assumes an isotropically escaping gas at a constant velocity originating from the nucleus. We adopt a mean expansion speed of $0.8 \times r_h^{-0.6}$~km~s$^{-1}$, following \citet{Biver:1999}, where $r_h$ is the heliocentric distance in au. The derived production rates are 
$\mathrm{Q_{CN}} =(1.6\pm0.5)\times10^{26}$~molecules~s$^{-1}$, 
$\mathrm{Q_{C_2}}=(8.8\pm0.8)\times10^{25}$~molecules~s$^{-1}$, and 
$\mathrm{Q_{C_3}}=(5.0\pm0.4)\times10^{24}$~molecules~s$^{-1}$. The logarithmic ratio between $\mathrm{Q_{C_2}}$ and $\mathrm{Q_{CN}}$ is $-0.26\pm0.14$, and between $\mathrm{Q_{C_3}}$ and $\mathrm{Q_{CN}}$ is $-1.51\pm0.14$. Although still C-chain depleted, this ratio is higher than the pre-perihelion upper limit ($\log\left(\mathrm{Q_{C_2}}/\mathrm{Q_{CN}}\right) \leq -0.8$, \citealp{SalazarManzano25}). 

Subsequent production rates from TRAPPIST observations at $r_h=1.76$~au find the activity decreases as 3I/ATLAS moves farther away from the Sun \citep[$\mathrm{Q_{CN}}  = (5.5\pm0.2)\times10^{25}$~molecules~s$^{-1}$;][]{jehin2011trappist, Jehin25_ATel}, and even later TRAPPIST measurements \citep{Jehin25_ATel_2} find a $\sim$50\% decrease in production rate from $r_h=1.76$~au to $r_h=1.98$~au, a trend that is consistent with our measurements despite the difference in profile and \citet{Haser:1957} model assumptions. 

We show the post-perihelion production rates reported to date in Figure \ref{fig:prod_rate_evo}.  We also fit the post-perihelion production rate evolution with a simple power law defined as $C\times \left(r_h\right)^n$, where $C$ is a constant and $r_h$ is the heliocentric distance in au. CN and $\mathrm{C_2}$ have similar power law indices, whereas $\mathrm{C_3}$ is nearly constant. A positive or flat heliocentric distance slope is not seen in any of the 85 comets in \citet{A'Hearn:1995}, suggesting this trend for $\mathrm{C_3}$ in 3I/ATLAS may arise from a non-physical driver, such as different assumptions in the computation of production rate (although we note that if so, the CN and $\mathrm{C}_2$ trends do not have this issue).

At larger heliocentric distances, CN production is commonly attributed to photolysis of HCN, although CN can also receive contributions from the degradation of complex refractory organic material on dust grains \citep{A'Hearn:1995, Fray2005}. The steep decline in CN production is therefore consistent with dominance by a near-nucleus volatile source whose release decreases rapidly as insolation drops. The shallower heliocentric dependence observed for $\mathrm{C}_2$ suggests a larger contribution from a distributed grain source. In contrast, the near-zero slope for $\mathrm{C}_3$ may indicate that its production is dominated by a distributed source whose contribution is delayed post-perihelion. However, if such a distributed source violates the assumptions of the Haser model, the derived production rates may acquire a spurious heliocentric dependence \citep{Cochran1985}.

We interpret the Fe\,\textsc{i} and Ni\,\textsc{i} emission lines using a simplified three-level atom approximation consisting of the ground state, a metastable lower level, and an excited upper level populated by resonance fluorescence under diluted solar radiation \citep{Preston:1967, Arpigny:1978}. Under statistical equilibrium and negligible collisional effects, the line intensities follow a Boltzmann--type relation: $\log_{10}(I\,\lambda^{3}/g f) = -\theta\,\chi_u + C$ with $\theta = 5040/T$, where $I$ is the line intensity integrated over the observed profile, $\lambda$ is the transition wavelength, $g$ and $f$ are the statistical weight of the lower level and the oscillator strength, $\chi_u$ is the excitation energy of the upper level in eV, and $T$ is the excitation temperature. The excitation temperature is empirically derived by fitting multiple Fe\,\textsc{i} lines spanning a range of upper--level energies. The fitted constant $C$ is then used to derive column densities and production rates accounting for solar dilution, and the partition function, following the formalism of \cite{Manfroid2021}. Consistent with their findings, we adopt $T(\mathrm{Ni\,\textsc{i}}) = T(\mathrm{Fe\,\textsc{i}}) + 180~\mathrm{K}$. Atomic parameters for Fe and Ni transitions are taken from the NIST Atomic Spectra Database \citep{Kramida2024}.

The derived Fe and Ni production rates are 
$\mathrm{Q_{Fe}} = (9.55\pm3.96)\times10^{25}$~atoms~s$^{-1}$, and
$\mathrm{Q_{Ni}} = (6.61\pm2.74)\times10^{25}$~atoms~s$^{-1}$. The logged ratio is $\log(\mathrm{Q_{Ni}} / \mathrm{Q_{Fe}}) = -0.16 ± 0.03$, which is more Fe-rich than pre-perihelion ratios from \citet{Hutsemekers25} ($\log(\mathrm{Q_{Ni}} / \mathrm{Q_{Fe}}) = 0.60\pm0.04$) though not as much so as the solar abundance of $\log(\mathrm{Ni/Fe})_\odot=-1.25\pm0.04$ \citep{Asplund09}.  

\begin{deluxetable}{cccc}
\tablenum{1}
\tablecaption{Best-fit values for the production rate power law fits in Figure \ref{fig:prod_rate_evo}. $A$ is a constant and $n$ is the power-law index.}\label{tab:prod_rates}
\tablewidth{\linewidth}
\tablehead{  \colhead{Species} &  \colhead{$A$}     & \colhead{$n$}\\
             \colhead{}        &  \colhead{s$^{-1}$}        & \colhead{} 
          }
\startdata
CN  &   $(4.6\pm1.7)\times10^{26}$   &   $-3.72\pm0.62$ \\
$\mathrm{C}_2$  &   $(2.8\pm0.7)\times10^{26}$   &   $-2.69\pm0.47$ \\
$\mathrm{C}_3$  &   $(4.5\pm1.8)\times10^{24}$   &   $+0.24\pm0.88$ 
\enddata
\end{deluxetable}

\subsection{Radial Distributions}\label{sec:radial}

\begin{figure*}
\includegraphics[width=\textwidth]{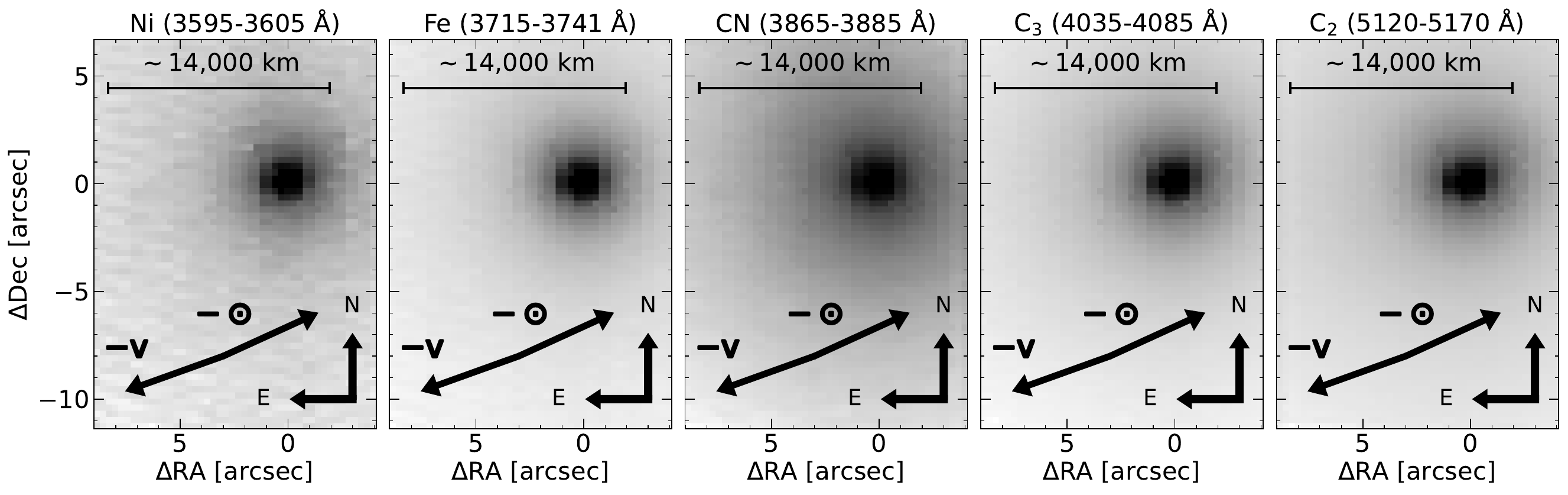}
\caption{Comparison of narrow-band images from the KCWI data cube in the spectral regions corresponding to Ni, Fe, CN, $\mathrm{C_3}$, and $\mathrm{C_2}$. The maximum value of each panel is set individually.}
\label{fig:KCWI_whitelight_comp}
\end{figure*}

Our IFU data enable spectro-spatial analyses of the emission species in 3I/ATLAS. At $\Delta=2.089$~au, the entire KCWI field of view corresponds to a projected physical size of $\sim30\,000\times25\,000$~km. 3I/ATLAS filled the entire medium slicer in acquisition images. Figure \ref{fig:KCWI_whitelight_comp} shows the KCWI 2D narrow-band images for spectral regions dominated by each observed feature: Ni, Fe, CN, $\mathrm{C}_3$, and $\mathrm{C}_2$. The CN emission is the most extensive. 

\begin{figure}
\includegraphics[width=\linewidth]{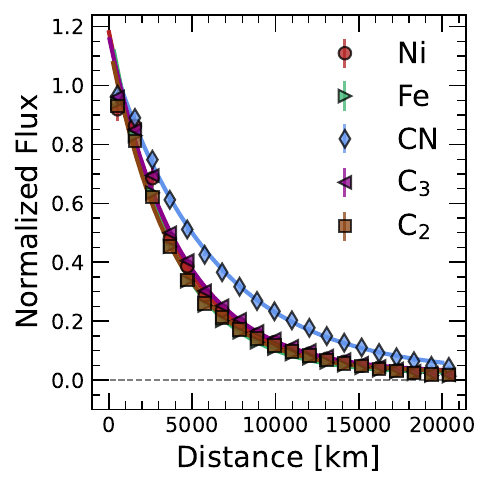}
\caption{Measured radial profiles of Ni, Fe, CN, $\mathrm{C}_3$ and $\mathrm{C}_2$ in 3I/ATLAS measured using the images shown in Figure \ref{fig:KCWI_whitelight_comp}. The maximum unbinned flux values for each feature are normalized to 1 to correct for luminosity differences. Each point denotes radially binned data with a radius step size of 1046~km. Exponential fits are overplotted.} 
\label{fig:KCWI_profiles}
\end{figure}

We next fit the radial profiles of Ni, Fe, CN, $\mathrm{C}_3$, and $\mathrm{C}_2$ with the same exponential decay model as \citet{Hoogendam25_KCWI}, which is defined as 
\begin{equation}
    A\times\exp\left[\frac{-x}{\tau}\right] + C,    
\end{equation}
where $\tau$ is the characteristic $e$-folding length scale of the radial profile and $A$ and $C$ are nuisance parameters. To account for differences in continuum level, we normalize the data to the maximum flux value. We use sigma-clipping with ten fit iterations to remove outliers. 

The apparent spatial extent of an emission depends on where the emitting species is produced, either directly from nucleus-released volatiles or from a distributed grain source in the coma, and on how long the species survives before being destroyed by sunlight. CN is efficiently produced from a volatile parent and survives a long time as a radical, giving it a large daughter scale length and a bright profile that persists far down the tailward coma. For $\mathrm{C}_2$ and $\mathrm{C}_3$ there may be distributed sources (e.g., organic-rich grains) and multi-step chemistry. This tends to make their spatial profiles different from CN and often less extended in the outer coma. Thus, it is unsurprising that CN has a longer $e$-folding length scale than the other species observed in 3I/ATLAS. 

\begin{deluxetable}{cccc}
\tablenum{2}
\tablecaption{Best fit values for the radial profiles in Figure \ref{fig:KCWI_profiles}.}\label{tab:spec}
\tablewidth{\linewidth}
\tablehead{  \colhead{Species} &  \colhead{$A$}     & \colhead{$\tau$}  & \colhead{$C$} \\
             \colhead{}        &  \colhead{}        & \colhead{[km]}      & \colhead{[km]} 
          }
\startdata
Ni  &   $1.112\pm0.009$   &   $3880.2\pm39.3$ & $0.110\pm0.001$ \\
Fe  &   $1.105\pm0.009$   &   $3694.9\pm36.3$ & $0.092\pm0.001$ \\
CN  &   $0.748\pm0.004$   &   $6053.1\pm67.6$ & $0.333\pm0.002$ \\
$\mathrm{C}_3$  &   $1.028\pm0.008$   &   $4193.9\pm44.9$ & $0.128\pm0.002$ \\
$\mathrm{C}_2$  &   $0.994\pm0.010$   &   $3833.4\pm44.8$ & $0.141\pm0.001$ 
\enddata
\end{deluxetable}

Figure \ref{fig:KCWI_profiles} shows our fits to the azimuthally averaged flux as a function of the physical radial distance for each emission feature; the best-fit parameters of the fits are given in Table \ref{tab:spec}. The CN flux is far more dispersed than the other species, with a characteristic $e$-folding lengthscale $\sim2000$~km longer. This lengthscale increased by a factor of $\sim$7 from the pre-perihelion measurement of $841.0\pm15.4$~km from \citet{Hoogendam25_KCWI}. In comparison, the Ni $e$-folding lengthscale increased by a factor of $\sim$6.5. 

Speculatively, the increase in $e$-folding radius is not what is expected from heliocentric-distance scaling, indicating a possible change in the dominant source and/or kinematics of the radicals. Enhanced, post-perihelion activity, potentially driven by thermal lag, could produce more distributed sources through increased grain release or fragmentation. Additionally, seasonal effects associated with a potential high-obliquity rotation pole may modify the illumination of active regions. Together, stored heat near perihelion and changing seasonal illumination could enhance dust and gas release after perihelion, broadening the coma and increasing the role of distributed sources.

\subsection{Symmetry and Jets}

\begin{figure*}
\includegraphics[width=\textwidth]{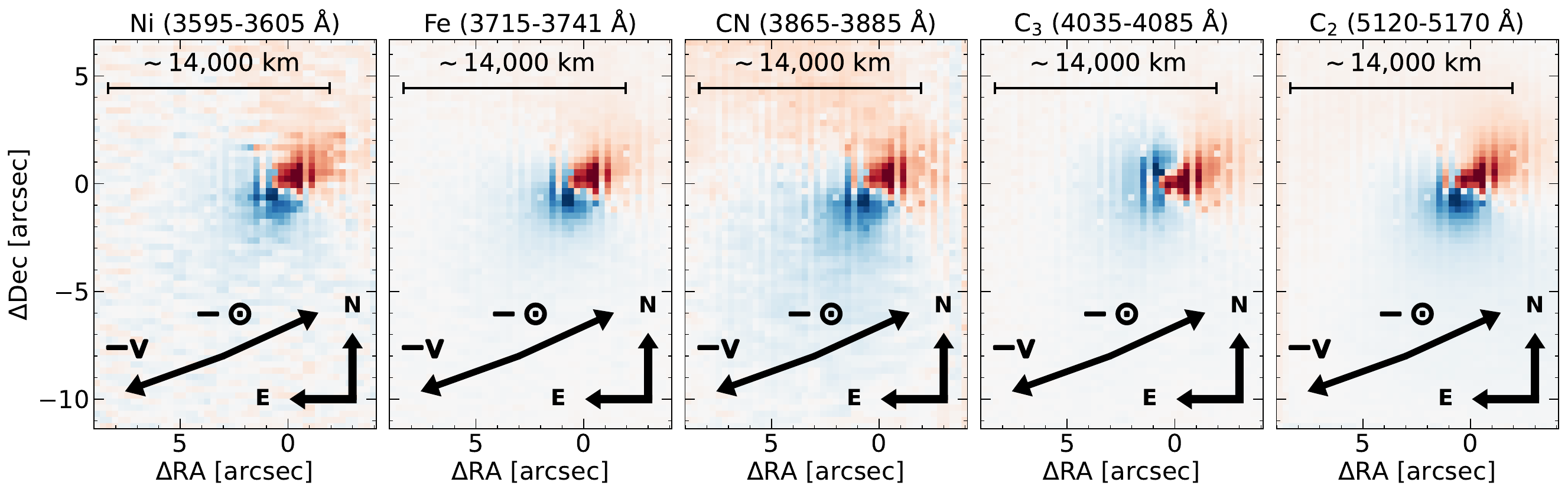}
\caption{Comparison of the non-symmetric residuals from the KCWI data cube for Ni, Fe, CN, $\mathrm{C_3}$ and $\mathrm{C_2}$. Each image from Figure \ref{fig:KCWI_whitelight_comp} is fit to determine the azimuthally averaged profile, which is then subtracted from the data. The resulting residuals show excess flux (red) and oversubtracted flux (blue). This excess flux corresponds to physical cometary emission (i.e., tails or jets), whereas the oversubtractions arise in the anti-tail direction.}
\label{fig:KCWI_whitelight_residual_comp}
\end{figure*}

Following \citet{Hoogendam25_KCWI}, we construct an azimuthally symmetric model of the comet flux and subtract it from the white light images. The angular flux residual profile for each emission feature in 3I/ATLAS is shown in Figure \ref{fig:KCWI_whitelight_residual_comp}. The model-subtracted image reveals excess flux in the anti-solar direction, consistent with a cometary tail. The residual is oversubtracted (by virtue of the spherical symmetry assumption) in the solar direction.

Interestingly, while the Ni, Fe, CN, and $\mathrm{C_2}$ features are roughly aligned with the Sun (the anti-solar direction is 294.5\degr East of North), $\mathrm{C_3}$ is misaligned with the anti-solar direction and the other emission features. One explanation could be that 3I/ATLAS, like Solar System comets, has an inhomogeneous composition, and the $\mathrm{C}_3$ activity originates from a region enriched in longer-chain hydrocarbons in the surface that feeds a jet. For example, if the parent species of $\mathrm{C}_2$ and $\mathrm{C}_3$ are $\mathrm{C}_2\mathrm{H}_2$, $\mathrm{C}_2\mathrm{H}_6$, and $\mathrm{C}_3\mathrm{H}_4$ \citep[e.g.,][]{Helbert05}, it may be the case that the $\mathrm{C}_3\mathrm{H}_4$ distribution is more localized to one region of the comet than those of $\mathrm{C}_2\mathrm{H}_2$ and $\mathrm{C}_2\mathrm{H}_6$, which could be more ubiquitous on the comet surface. 

Alternatively, the reported sunward jet in pre-perihelion observations \citep[e.g.,][]{Serra-Ricart26} could potentially still be active and now pointing away from the Sun. If so, it may be possible that the pre-perihelion activity depleted that specific part of 3I/ATLAS of most of its volatiles, but heavier hydrocarbons remain and drive the $\mathrm{C}_3$ activity in particular. This contrasts with other regions that are less sublimation-depleted, which would still emit bulk volatiles that are swept into an anti-solar tail. One challenge to this scenario is that the hydrocarbon parent species of $\mathrm{C}_2$ do not seem to have the same dependence despite also originating from similarly heavy hydrocarbons.
 
Figure \ref{fig:KCWI_whitelight_residual_comp} also shows that the $\mathrm{C}_3$ emission magnitude is less symmetrical around $\theta\approx270$~degrees from North through East than the other features. This is likely due to additional flux contributions from $\mathrm{C}_2$, which overlaps with the $\mathrm{C}_3$ region (we do not see this effect for $\mathrm{C}_2$, because our $\mathrm{C}_2$ wavelength range excludes $\mathrm{C}_3$ lines).

\section{Remarks on the Composition of 3I/ATLAS}
\begin{figure*}
\includegraphics[width=\textwidth]{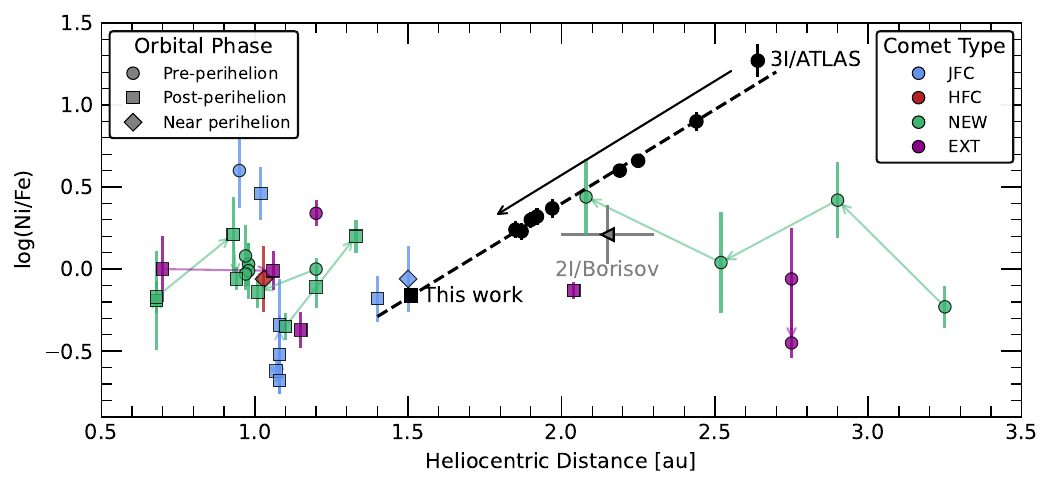}
\caption{Ni/Fe ratio evolution for 3I/ATLAS (black circles; \citealp{Hutsemekers25} and this work, black square) compared to 2I/Borisov \citep{Opitom2021} and Solar System comets \citep{Manfroid2021}. Comets connected with lines are repeat observations of the same comet, and the arrows denote the direction of movement. 3I/ATLAS initially had a higher Ni/Fe ratio than any other observed comet, but after its perihelion passage, it is similar to the Solar System comet 9P/Tempel 1. Comet classification follows \citet{Manfroid2021}: JFCs are Jupiter-family comets, HFCs are Halley-family comets, NEW comets are dynamically new ($a < 10\,000$~au), and EXT are directly from the Oort cloud ($a > 10\,000$~au).}
\label{fig:Ni/Fe_ratio_evol}
\end{figure*}

Pre-perihelion observations of Ni \citep{Rahatgaonkar2025, Hoogendam25_KCWI, Hoogendam25_SNIFS} and Fe \citep{Hutsemekers25} emission reveal that 3I/ATLAS is an outlier among Solar System comets in its Ni/Fe ratio, both compared to objects at distances of 2--3~au and the entire sample from \citet{Manfroid2021} regardless of $r_h$. Figure \ref{fig:Ni/Fe_ratio_evol} shows the evolution of the logged Ni/Fe ratio for the \citet{Manfroid2021} sample of Solar System comets as a function of $r_h$. Unfortunately, no Solar System comet is as well-measured or as comprehensively observed as 3I/ATLAS.

3I/ATLAS shows strong, linear $\log\left(\mathrm{Ni/Fe}\right)$ evolution as a function of $r_h$. We fit a line to the data and find a slope of $1.15\pm0.05$ (and an intercept of $-1.89\pm0.10$). 3I/ATLAS has $\log\left(\mathrm{Ni/Fe}\right)\approx1.4$ at $\sim$2.5 au, whereas the only two Solar System comets (both from the Oort cloud, \citealp{Oort1950}) measured at comparable distance have values consistent with 0, suggesting either that 3I/ATLAS is extraordinarily enriched in Ni or the Solar System comparison sample does not capture the full diversity of cometary compositions and behaviors at that distance. 3I's evolution from an extreme outlier to outgassing Ni/Fe consistent with Solar System comets suggests that the latter is the most likely explanation. 

With a perihelion distance of 1.36~au, 3I/ATLAS does not approach the Sun as closely as many of the comets from \citet{Manfroid2021}. Few Solar System comets have more than one $\log\left(\mathrm{Ni/Fe}\right)$ measurement, but those that do tend not to show the significant linear evolution seen in 3I/ATLAS. However, where changes have been observed at $r_h <$ 2 au, they have been in the same direction: Ni/Fe generally increasing with $r_h$ \citep[see also][]{Rahatgaonkar2025, Hutsemekers25}. 

\section{Conclusions}
We present the first post-perihelion IFU analysis of 3I/ATLAS. At the time of observation, 3I/ATLAS was at $r_h=1.51$~au. We observe CN, Ni, and Fe activity, which were also seen in pre-perihelion observations \citep[e.g.,][]{Rahatgaonkar2025, Hoogendam25_KCWI, SalazarManzano25, Hoogendam25_SNIFS, Hutsemekers25}. Additionally, we report $\mathrm{C}_2$, $\mathrm{C}_3$, and CH activity. The CN emission has a larger $e$-folding distance than the other species by a factor of $\sim$1.5, consistent with CN having a longer survival time as a radical. 

We find evidence that the $\mathrm{C}_3$ emission in particular may trace a jet emerging at a different angle than the other species. The jet may arise from either inhomogeneities in the surface composition of 3I/ATLAS or from continuation of the previously reported jets \citep[e.g.,][]{Serra-Ricart26}. Further post-perihelion IFU data will improve the understanding of this feature. 

3I/ATLAS has uniquely comprehensive pre- and post-perihelion observations that track the evolution of atomic Ni and Fe---something only one Solar System comet, C/2002 V1 (NEAT), has. Even if the driving mechanism for Ni emission in the solar system differs from that of interstellar comets, Ni and Fe measurements for interstellar comets offer a promising pathway to explore the primordial metallicity of other planetary systems. Continued post-perihelion observations will provide further insights into the post-perihelion evolution of the Ni/Fe ratio and how Fe and Ni emission may eventually ``turn off'' with increasing heliocentric distance. 

The Legacy Survey of Space and Time \citep{LSST_2019} will discover additional interstellar objects and, crucially, may detect many of them before their perihelion passage, enabling improved understanding of the solar radiation processing a perihelion passage induces on this type of object. Increasing the sample of interstellar and Solar System comets with detections of Ni and Fe will enable further population-level studies to understand the metal content of interstellar objects and the primordial planetary formation material in other solar systems.

\begin{acknowledgments}
W.B.H. acknowledges support from the National Science Foundation Graduate Research Fellowship Program under Grant No. 2236415. 

D.O.J. acknowledges support from NSF grants AST-2407632, AST-2429450, and AST-2510993, NASA grant 80NSSC24M0023, and HST/JWST grants HST-GO-17128.028 and JWST-GO-05324.031, awarded by the Space Telescope Science Institute (STScI), which is operated by the Association of Universities for Research in Astronomy, Inc., for NASA, under contract NAS5-26555.

The Shappee group at the University of Hawai\okina i at M\={a}noa is supported by NSF (grant AST-2407205) and NASA (grants HST-GO-17087, 80NSSC24K0521, 80NSSC24K0490, 80NSSC23K1431).

K.J.M., J.J.W., and A.H.\ acknowledge support from the Simons Foundation through SFI-PD-Pivot Mentor-00009672. 
J.T.H. acknowledges support from NASA through the NASA Hubble Fellowship grant HST-HF2-51577.001-A, awarded by STScI. STScI is operated by the Association of Universities for Research in Astronomy, Incorporated, under NASA contract NAS5-26555. 

C.A. and K.M. acknowledge support from NASA grants JWST-GO-02114, JWST-GO-02122, JWST-GO-03726, JWST-GO-04217, JWST-GO-04436, JWST-GO-04522, JWST-GO-05057, JWST-GO-05290, JWST-GO-06023, JWST-GO-06213, JWST-GO-06583, and JWST-GO-06677. Support for these programs was provided by NASA through a grant from the Space Telescope Science Institute, which is operated by the Association of Universities for Research in Astronomy, Inc., under NASA contract NAS5-03127.

Some of the data presented herein were obtained at Keck Observatory, which is a private 501(c)3 non-profit organization operated as a scientific partnership among the California Institute of Technology, the University of California, and the National Aeronautics and Space Administration. The Observatory was made possible by the generous financial support of the W. M. Keck Foundation.

This research has made use of the Keck Observatory Archive (KOA), which is operated by the W. M. Keck Observatory and the NASA Exoplanet Science Institute (NExScI), under contract with the National Aeronautics and Space Administration.

The authors wish to recognize and acknowledge the very significant cultural role and reverence that the summit of Maunakea has always had within the Native Hawaiian community. We are most fortunate to have the opportunity to conduct observations from this mountain.

This research made use of \texttt{PypeIt}\footnote{\url{https://pypeit.readthedocs.io/en/latest/}}, a Python package for semi-automated reduction of astronomical data \citep{pypeit:joss_pub, pypeit:zenodo}.

\end{acknowledgments}

\facility{Keck:II (KCWI)}

\bibliography{sample701}{}
\bibliographystyle{aasjournal}

\end{document}